\documentclass[onecolumn,showpacs,aps]{revtex4}

\usepackage{graphicx}
\usepackage{dcolumn}
\usepackage{amsmath}
\usepackage{epsfig}
\usepackage{pstricks}

\input pubboard/babarsym

\def\ap{\ensuremath{A_{\parallel} }\xspace}
\def\az{\ensuremath{A_{0} }\xspace}
\def\at{\ensuremath{A_{\perp} }\xspace}
\def\thetakstar{\ensuremath{\theta_{K^*}}\xspace}
\def\thetatr{\ensuremath{\theta_{tr}}\xspace}
\def\phitr{\ensuremath{\varphi_{tr}}\xspace}
\def\ctwob{\ensuremath{\cos\! 2 \beta   }\xspace}

\newcommand{\BABARPubYear}    {02}
\newcommand{\BABARPubNumber}  {01}

\newcommand{\SLACPubNumber} {9153}

\def\figurebox#1#2#3{%
    \def\arg{#3}%
    \ifx\arg\empty
    {\hfill\vbox{\hsize#2\hrule\hbox to #2{\vrule\hfill\vbox to
          #1{\hsize#2\vfill}\vrule}\hrule}\hfill}%
    \else
    {\hfill\epsfbox{#3}\hfill}%
    \fi}

\long\def\inst#1{\par\nobreak\kern 4pt\nobreak
    {\it #1}\par\vskip 10pt plus 3pt minus 3pt}

\begin{document}

\preprint{\babar-PUB-\BABARPubYear/\BABARPubNumber} 
\preprint{SLAC-PUB-\SLACPubNumber} 

\begin{flushright}
\babar-CONF-\BABARPubYear/\BABARPubNumber\\
SLAC-PUB-\SLACPubNumber\\
March, 2002 
\end{flushright}

\title{
\vskip 10mm
{\large \bf
 Improved Measurement of the {\boldmath\CP}-violating Asymmetry
 Amplitude {\boldmath\stwob} 
}
\begin{center} 
\vskip 10mm
The \babar\ Collaboration
\end{center}
}

\date{\today} 

\begin{abstract}
We present updated results on time-dependent \CP-violating
asymmetries in neutral $B$ decays to several \CP eigenstates. The
measurements use a data sample of about 62 million $\FourS\to B\Bbar$
decays collected between 1999 and 2001 by the \babar\ detector at the
\pep2\ asymmetric-energy \BF\ at SLAC. In this sample we study events 
in which one neutral $B$ meson is fully reconstructed in a final state
containing a charmonium meson and the flavor of the other neutral $B$
meson is determined from its decay products. The amplitude of the
\CP-violating asymmetry, which in the Standard Model is proportional to
\stwob, is derived from the decay time distributions in such events. We
measure $\stwob = 0.75 \pm 0.09\ \stat \pm 0.04\ \syst$ and 
$|\lambda| = 0.92 \pm 0.06\ \stat \pm 0.02\ \syst$. The latter is
consistent with the Standard Model expectation of no direct \CP
violation. These results are preliminary.  
\end{abstract}

\pacs{13.25.Hw, 12.15.Hh, 11.30.Er}

\collaboration{The \babar\ Collaboration}
%
\author{B.~Aubert}
\author{D.~Boutigny}
\author{J.-M.~Gaillard}
\author{A.~Hicheur}
\author{Y.~Karyotakis}
\author{J.~P.~Lees}
\author{P.~Robbe}
\author{V.~Tisserand}
\author{A.~Zghiche}
\affiliation{Laboratoire de Physique des Particules, F-74941 Annecy-le-Vieux, France }
\author{A.~Palano}
\author{A.~Pompili}
\affiliation{Universit\`a di Bari, Dipartimento di Fisica and INFN, I-70126 Bari, Italy }
% ...
\collaboration{The \babar\ Collaboration}
\noaffiliation

\noaffiliation

\maketitle
The Standard Model of electroweak interactions describes \CP violation
in weak decays as a consequence of a complex phase in the
three-generation Cabibbo-Kobayashi-Maskawa~\cite{CKM} (CKM) quark-mixing
matrix. In this picture, measurements of \CP -violating asymmetries in 
the time distributions of \Bz decays to charmonium final states provide
a direct measurement of \stwob~\cite{BCP}, where $\beta \equiv \arg \left[\,
  -V_{\rm cd}^{}V_{\rm cb}^* / V_{\rm td}^{}V_{\rm tb}^*\, \right]$. 

Measurements of the \CP -violating asymmetry parameter \stwob\ have
recently been published by the \babar~\cite{babar-stwob-prl} and
Belle~\cite{belle-stwob-prl} collaborations from data taken between 1999 
and summer 2001 at the PEP-II and KEKB asymmetric-energy \epem colliders. 
These results, which followed less precise
measurements~\cite{OPALCDFALEPH}, established \CP violation in the \Bz  
system. In this paper, we report an updated measurement of \stwob ,
using a sample of 62 million \Bz decays collected with the
\babar\ detector. Since our previous measurement,
we have added a sample of 30 million \Bz decays collected
in the latter half of 2001, and have improved
data reconstruction and analysis techniques. The measurement technique
is described in detail in Ref.~\cite{babar-stwob-prd}. The discussion
here is limited to the changes in the analysis with respect to the
published results~\cite{babar-stwob-prl,babar-stwob-prd}. 

Since the \babar\ detector is described in detail
elsewhere~\cite{babar-detector-nim},  
only a brief description is given here. 
Surrounding the beam-pipe is a silicon vertex tracker (SVT), which
provides precise measurements of the trajectories of charged particles
as they leave the \epem interaction point.  
Outside of the SVT, a 40-layer drift chamber (DCH) 
allows measurements of track momenta in a 1.5\,T magnetic field as well as
energy-loss measurements, which contribute to charged
particle identification. Surrounding the DCH is a detector of internally
reflected Cherenkov radiation (DIRC), which provides charged hadron
identification. Outside of the DIRC is a CsI(Tl) electromagnetic
calorimeter (EMC) that is used to detect photons, provide electron
identification and reconstruct neutral hadrons. The EMC is surrounded by a
superconducting coil, which creates the magnetic field for momentum
measurements.  Outside of the coil, the 
flux return is instrumented with resistive plate chambers interspersed with 
iron (IFR) for the identification of muons and long-lived neutral hadrons.
We use the GEANT4~\cite{geant4} software to simulate interactions of particles
traversing the \babar\ detector.

From approximately 56\,\invfb\ of data recorded at the \FourS\ resonance, 
corresponding to 62 million produced \BB\ pairs, we reconstruct a
sample  of neutral $B$ mesons, $B_{\CP}$, decaying to the final states    
$\jpsi\KS\ (\KS \to \pipi,\ \ppz)$,
$\psitwos\KS\ (\KS \to \pipi)$, 
$\chicone\KS\ (\KS \to \pip\pim) $,
$\jpsi\Kstarz\ (\Kstarz\to \KS\piz,\ \KS\to\pipi)$, and
$\jpsi\KL$. 
The $\jpsi$ and $\psitwos$ mesons are reconstructed through their decays
to $e^+e^-$ and $\mu^+\mu^-$; the $\psitwos$ is also reconstructed
through its decay to $\jpsi\pi^+\pi^-$. The $\chicone$ meson is
reconstructed in the decay mode $\jpsi\gamma$. We examine each of the
events in the $B_{\CP}$ sample for evidence that the recoiling neutral
$B$ meson decayed as a \Bz or \Bzb (flavor tag).

The decay-time distribution of $B$ decays to a \CP eigenstate with a \Bz
or \Bzb tag can be expressed in terms of a complex parameter $\lambda$
that depends on both the \Bz-\Bzb oscillation amplitude and the amplitudes
describing \Bzb and \Bz decays to this final
state~\cite{lambda}. The decay rate  ${\rm f}_+({\rm f}_-)$ when the 
tagging meson is a $\Bz (\Bzb)$ is given by 

\begin{eqnarray}
{\rm f}_\pm(\, \deltat) = {\frac{{\rm e}^{{- \left| \deltat \right|}/\tau_{\Bz} }}{4\tau_{\Bz}
}}  \times  \left[ \ 1 \hbox to 0cm{}
\pm \frac{{2\mathop{\cal I\mkern -2.0mu\mit m}}
\lambda}{1+|\lambda|^2}  \sin{( \deltamd  \deltat )} 
\mp { \frac{1  - |\lambda|^2 } {1+|\lambda|^2} }  
  \cos{( \deltamd  \deltat) }   \right],
\label{eq:timedist}
\end{eqnarray}

\vskip12pt\noindent
where $\Delta t = t_{\rm rec} - t_{\rm tag}$ is the difference between 
the proper decay time of the reconstructed $B$ meson ($B_{\rm rec}$) and 
the proper decay time of the tagging $B$ meson ($B_{\rm tag}$),
$\tau_{\Bz}$ is the \Bz lifetime, \deltamd is the mass difference 
determined from \Bz-\Bzb oscillations, and the lifetime difference between
the neutral \B\ mass eigenstates is assumed to be negligible. The sine
term in Eq.~\ref{eq:timedist} is due to the interference between direct
decay and decay after flavor change, and the cosine term is due to the
interference between two or more decay amplitudes with different weak
phases. Evidence for \CP violation can be observed as a difference
between the \deltat distributions of \Bz- and \Bzb-tagged events or as
an asymmetry with respect to $\deltat = 0$ for either flavor tag. 
\par
In the Standard Model, $\lambda=\eta_f e^{-2i\beta}$ for
charmonium-containing $b\to\ccbar s$ decays where $\eta_f$ is the \CP 
eigenvalue of the final state $f$. Thus, the time-dependent
\CP-violating asymmetry is  

\begin{eqnarray}
A_{\CP}(\deltat) &\equiv&  \frac{ {\rm f}_+(\deltat)  -  {\rm f}_-(\deltat) }
{ {\rm f}_+(\deltat) + {\rm f}_-(\deltat) } = -\eta_f \stwob
\sin{ (\deltamd \, \deltat )} ,  
\label{eq:asymmetry}
\end{eqnarray}

\vskip12pt\noindent
with $\eta_f=-1$ for 
$\jpsi\KS$, 
$\psitwos\KS$, and
$\chicone \KS$, and
$+1$ for $\jpsi\KL$. 

The measurement of \stwob with the decay mode $B\to\jpsi\Kstarz
(\Kstarz\to\KS\piz)$ is experimentally complicated by the presence of both
even (L=0, 2) and odd (L=1) orbital angular momenta in the final state.
With the measured \CP-even and \CP-odd contributions to the decay
rate~\cite{BABARTRANS}, the experimental sensitivity to \stwob\ is
reduced by 24\% compared to pure \CP eigenstates. 
The interference between \CP-even and \CP-odd amplitudes in this mode 
allows a measurement of \ctwob up to a sign ambiguity.
The time- and angle-dependent decay rate ${\rm f}_{+} ({\rm f}_{-})$
when the tagging meson is a $\Bz (\Bzb)$ is given by 

\begin{eqnarray}
{\rm f}_{\pm}(\deltat,\vec{\omega}) &=& 
\frac{e^{-|\deltat|/\tau_{\Bz}}}{4\tau_{\Bz}} \bigg[ 
{I(\vec{\omega};\vec{\cal A})} \mp  \Big\{
{C(\vec{\omega};\vec{\cal A})}
\cos(\deltamd \deltat) \nonumber \\
& & +\Big[
{S_{\stwob}{(\vec{\omega};\vec{\cal A})}}\stwob  + 
{S_{\ctwob}{(\vec{\omega};\vec{\cal A})}}\ctwob \Big]
\sin(\deltamd \deltat) \Big\}\bigg] 
\label{eq:angle}
\end{eqnarray}

\vskip12pt\noindent
where the coefficients $I$, $C$, $S_{\stwob}$, and $S_{\ctwob}$ are
functions of three transversity angles $\vec{\omega}$ and the previously
measured transversity amplitudes $\vec{\cal A}$~\cite{BABARTRANS} (see
appendix~\ref{appendixa}).  

The event selection, lepton and charged kaon identification, and \jpsi
and $\psitwos$ reconstruction used in this analysis are similar to
those described in Ref.~\cite{babar-stwob-prl,babar-stwob-prd}.
Since these earlier publications, significant improvements have been made in
the analysis.
Charged kaon identification has improved due to a better 
alignment of the Cherenkov detector and better Cherenkov angle
reconstruction. 
For the $\Bz\to\jpsi\KL$ selection, we have loosened the muon selection
requirements for $\jpsi\to\mu^+\mu^-$ and the $\piz$ veto for \KL
candidates. 
In the  $\KS \to \pipi$ selection for  $\Bz\to\jpsi\KS$ candidates, the
requirement on the $\pip\pim$ mass has been relaxed to $472 <
m(\pip\pim) < 522$~MeV/$c^2$.
We have increased the sensitivity to \stwob for the mode $J/\psi K^{*0}
(K^{*0}\to K^0_S\pi^0)$ by taking into account the transversity
angles for each event instead of integrating out the angle dependence. 
Events reconstructed in this mode
that have a candidate in the mode $J/\psi K^{*+} (K^{*+}\to
K^0_S\pi^+)$ are rejected. 
In addition, the whole dataset has been processed with 
a uniform reconstruction algorithm and detector calibration.
This provides, in particular, better alignment
of the tracking system and improved track reconstruction efficiency for
the 20~fb$^{-1}$ of data collected in 1999--2000.  
For example, the event yield increased by 11\% (37\%) for the
$\eta_f=-1$ ($\jpsi\KL$) sample while the purity only decreased by 4\%
(3\%). The effect of all improvements decreases the error on 
\stwob\ scaled to the same integrated luminosity by 13\%. 
\par
Candidates in the $B_{\CP}$ sample are selected by requiring that the
difference $\Delta E$ between their energy and the beam energy in the 
center-of-mass frame be less than three standard deviations from zero. 
For modes involving \KS, the beam-energy substituted mass
$\mes=\sqrt{{(E^{\rm 
      cm}_{\rm beam})^2}-(p_B^{\rm cm})^2}$ must be greater than
$5.2\gevcc$. The resolution for $\Delta E$ is about 10\mev, except for
the $\KS\to\piz\piz$ (33\mev), the $\jpsi \Kstarz$ (20\mev) 
and the $\jpsi\KL$ (3.5\mev after $B$ mass constraint) modes.
For the purpose of determining numbers of events and purities, a signal
region $5.270\ (5.273) < \mes < 5.290\ (5.288)\gevcc$ is used for modes 
containing \KS ($\Kstarz$). The signal region for the mode $\jpsi\KL$
is defined by $|\Delta E| < 10\mev$.  
\par
A measurement of $A_{\CP}$ requires a determination of the experimental
$\Delta t$ resolution and the fraction of events in which the tag
assignment is incorrect. A mistag fraction $w$ reduces the observed
\CP asymmetry by a factor $(1-2w)$. 
Mistag fractions and $\Delta t$ resolution functions 
are determined from a sample $B_{\rm flav}$ 
of neutral $B$ decays to flavor eigenstates 
consisting of the channels 
$D^{(*)-}h^+ (h^+=\pi^+,\rho^+$, and $a_1^+)$ and $\jpsi\Kstarz
(\Kstarz\to\Kp\pim)$. 
Validation studies are performed with a control sample of charged $B$
mesons decaying to the final states $\jpsi K^{(*)+}$, $\psitwos
K^+$, $\chicone \Kp$, and 
$D^{(*)0}\pip$. 
\par
The methods for flavor tagging and vertex reconstruction, and the
determination of \deltat, are described in
Ref.~\cite{babar-stwob-prd}.     
For flavor tagging, we exploit information from the recoil $B$ decay in
the event. The charges of energetic 
electrons and muons from semileptonic $B$ decays, kaons, soft pions from
\Dstar decays, and  high momentum particles are correlated with
the flavor of the decaying $b$ quark. For example, a positive lepton
indicates a \Bz tag. 
About 68\% of the events can be assigned to one of four hierarchical,
mutually exclusive tagging categories. The remaining untagged events are
excluded from further analysis. 
\par 
For a lepton tag we require an electron or muon candidate with a
center-of-mass momentum $p_{\rm cm} >1.0$ or $1.1\gevc$,
respectively. This efficiently selects primary leptons from semileptonic
$B$ decays and reduces contamination due to oppositely-charged leptons from
charm decays. Events satisfying these criteria are assigned to the {\tt
  Lepton} category unless the lepton charge and the net charge of all
kaon candidates indicate opposite flavor tags. Events without a lepton
tag but with a non-zero net kaon charge are assigned to the {\tt Kaon}
category.  
All remaining events are passed to a neural network algorithm whose main
inputs are the momentum and charge of the track with the highest
center-of-mass momentum, and the outputs of secondary networks, trained
with Monte Carlo samples to identify primary leptons, kaons, and soft
pions. Based on the output of the neural network algorithm, events are
tagged as \Bz or \Bzb and assigned to the {\tt NT1} (more certain tags)
or {\tt NT2} (less certain tags) category, or not tagged at all.  
The tagging power of the {\tt NT1} and {\tt NT2} categories arises
primarily from soft pions and from recovering unidentified isolated
primary electrons and muons.  

The time interval \deltat between the two $B$ decays is calculated
from the measured separation \deltaz between the decay vertex of the 
reconstructed  $B$ meson ($B_{\rm rec}$) and the vertex of the
flavor-tagging $B$ meson ($B_{\rm tag}$) along the $z$ axis. 
The calculation of \deltat includes an event-by-event
correction for the direction of the $B_{\rm rec}$ with respect to  
the $z$ direction in the $\FourS$ frame. We determine the $z$ position
of the $B_{\rm rec}$ vertex from
the charged tracks that constitute the $B_{\rm rec}$ candidate. The
decay vertex of the $B_{\rm tag}$ is determined by fitting the tracks not
belonging to the $B_{\rm rec}$ candidate to a common vertex. An
additional constraint on the tagging vertex comes from a pseudotrack 
computed from the  $B_{\rm rec}$ vertex and three-momentum,
the beam-spot (with a vertical size of 10~\mum), and the \FourS
momentum. For 99.5\% of the reconstructed events the r.m.s. \deltaz
resolution is 180\mum. An accepted candidate must have converged fits
for the $B_{\rm rec}$ and $B_{\rm tag}$ vertices, a
\deltat\ error less than 2.5\ps, and a measured $\vert \deltat \vert <
20 \ps$. The fraction of events in data satisfying these requirements is
93\%. 
\par
\begin{figure}[!h]
\begin{center}%
\mbox{\epsfig{figure=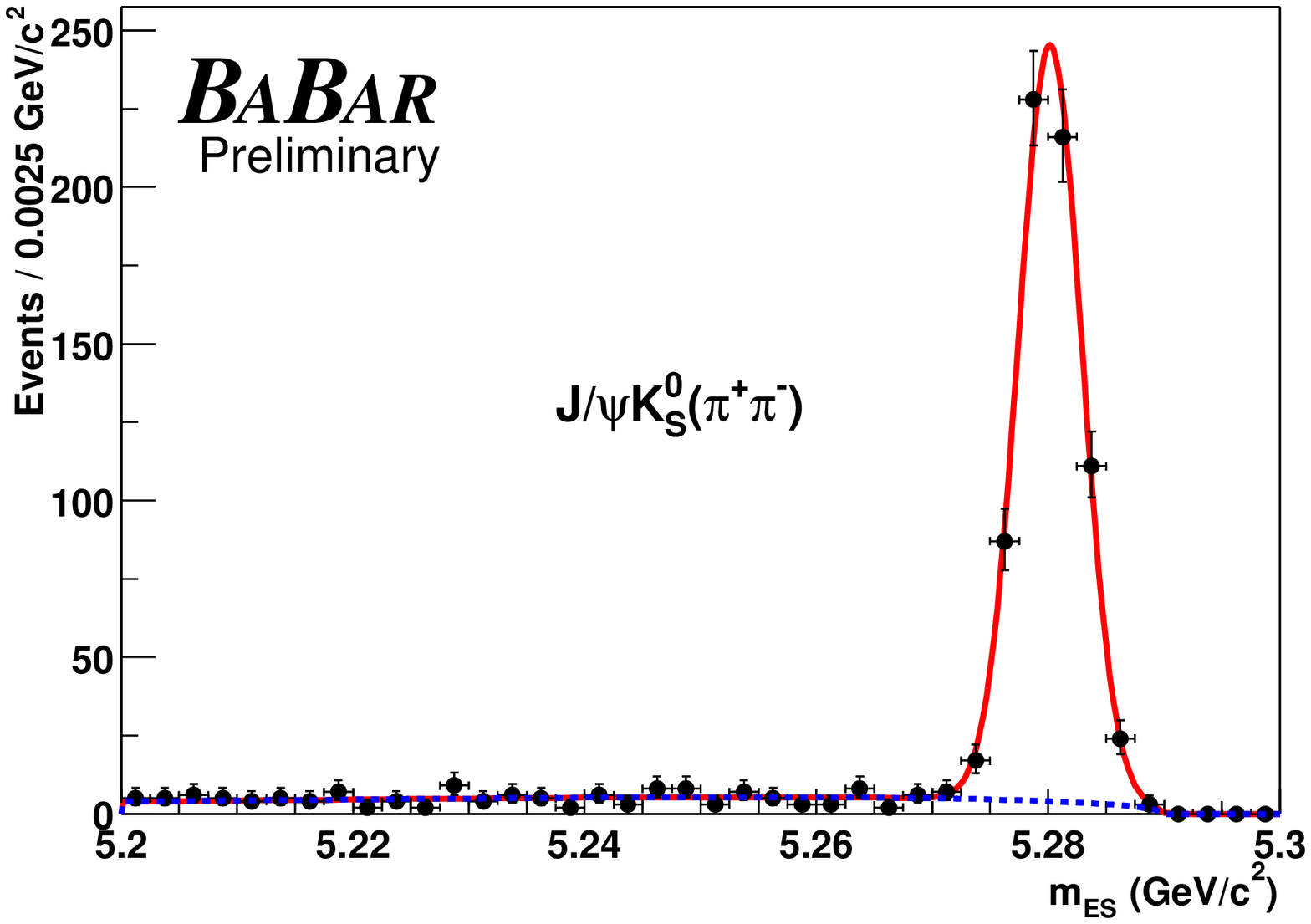,width=0.495\linewidth}
\epsfig{figure=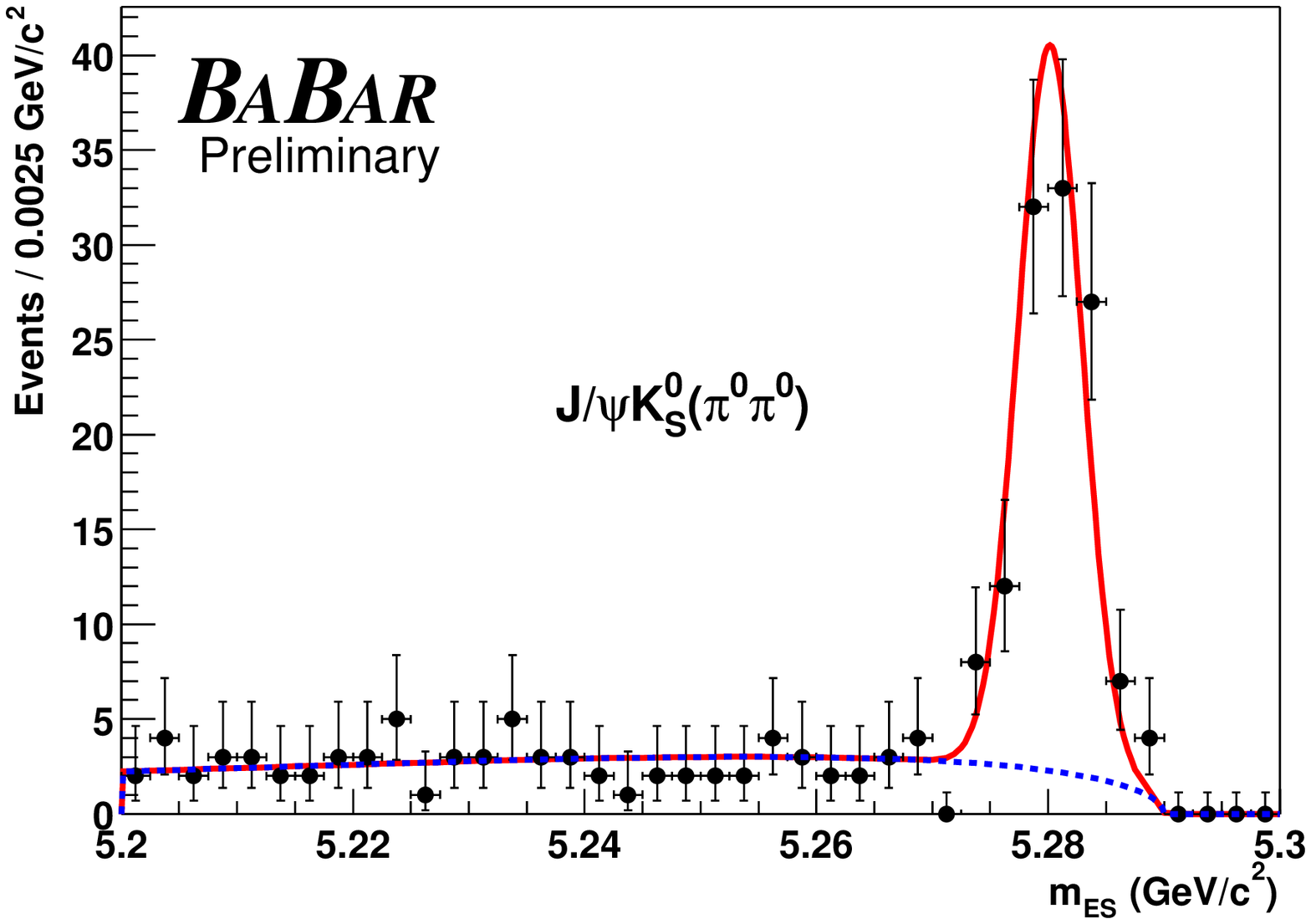,width=0.495\linewidth}
\put(-460,110){{\large a)}}
\put(-205,110){{\large b)}}}
\mbox{\epsfig{figure=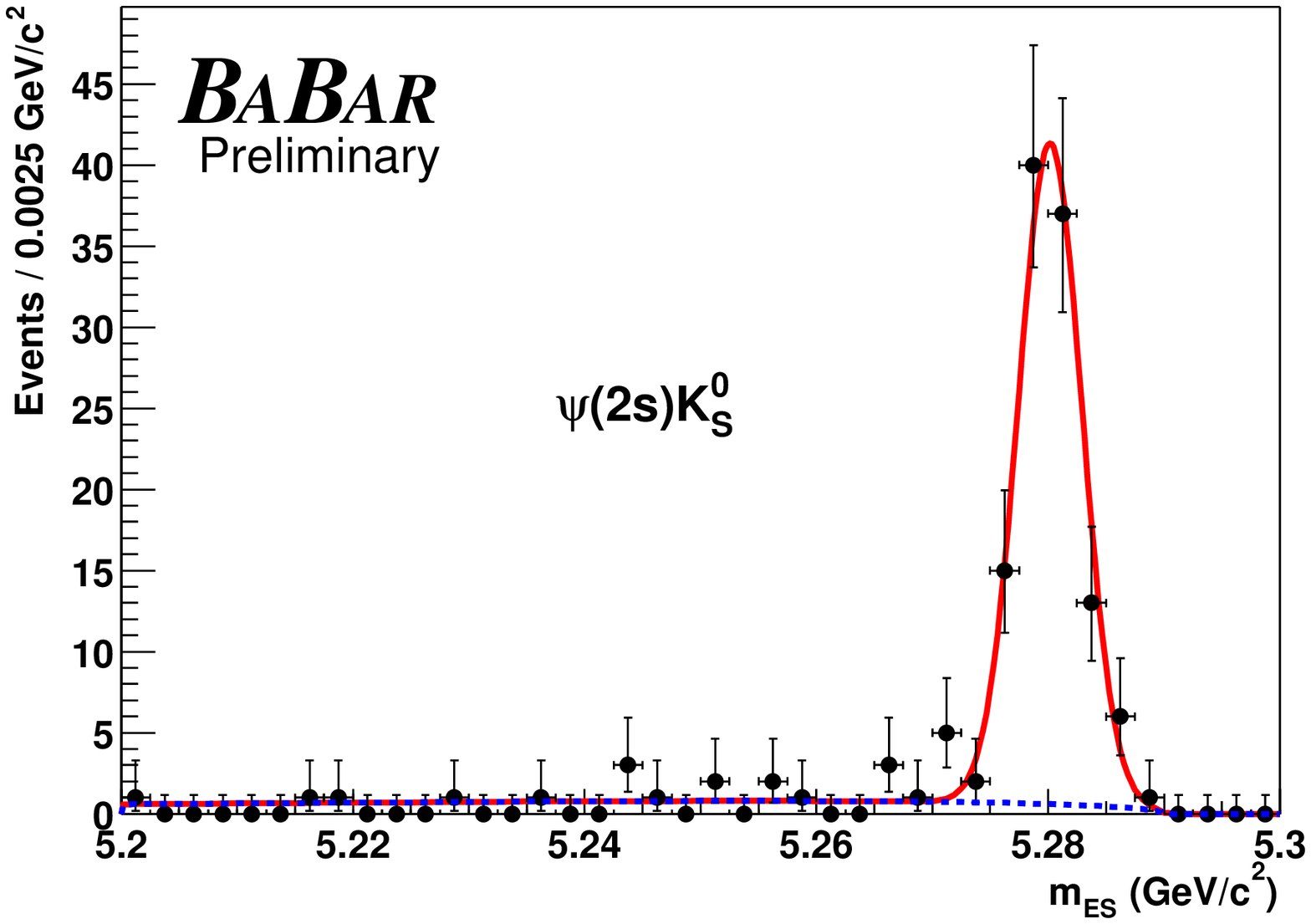,width=0.495\linewidth}
\epsfig{figure=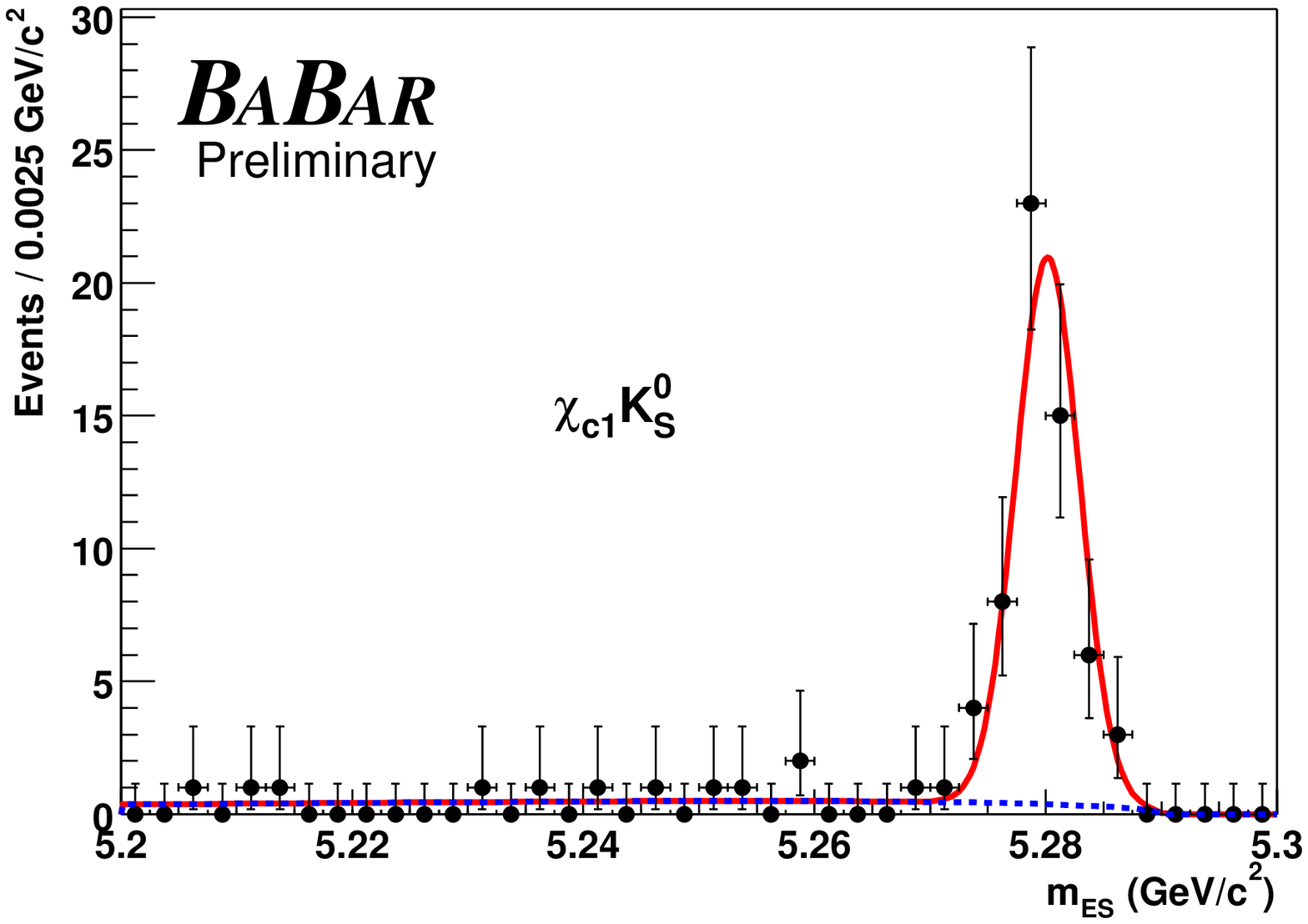,width=0.495\linewidth}
\put(-460,110){{\large c)}}
\put(-205,110){{\large d)}}}
\mbox{\epsfig{figure=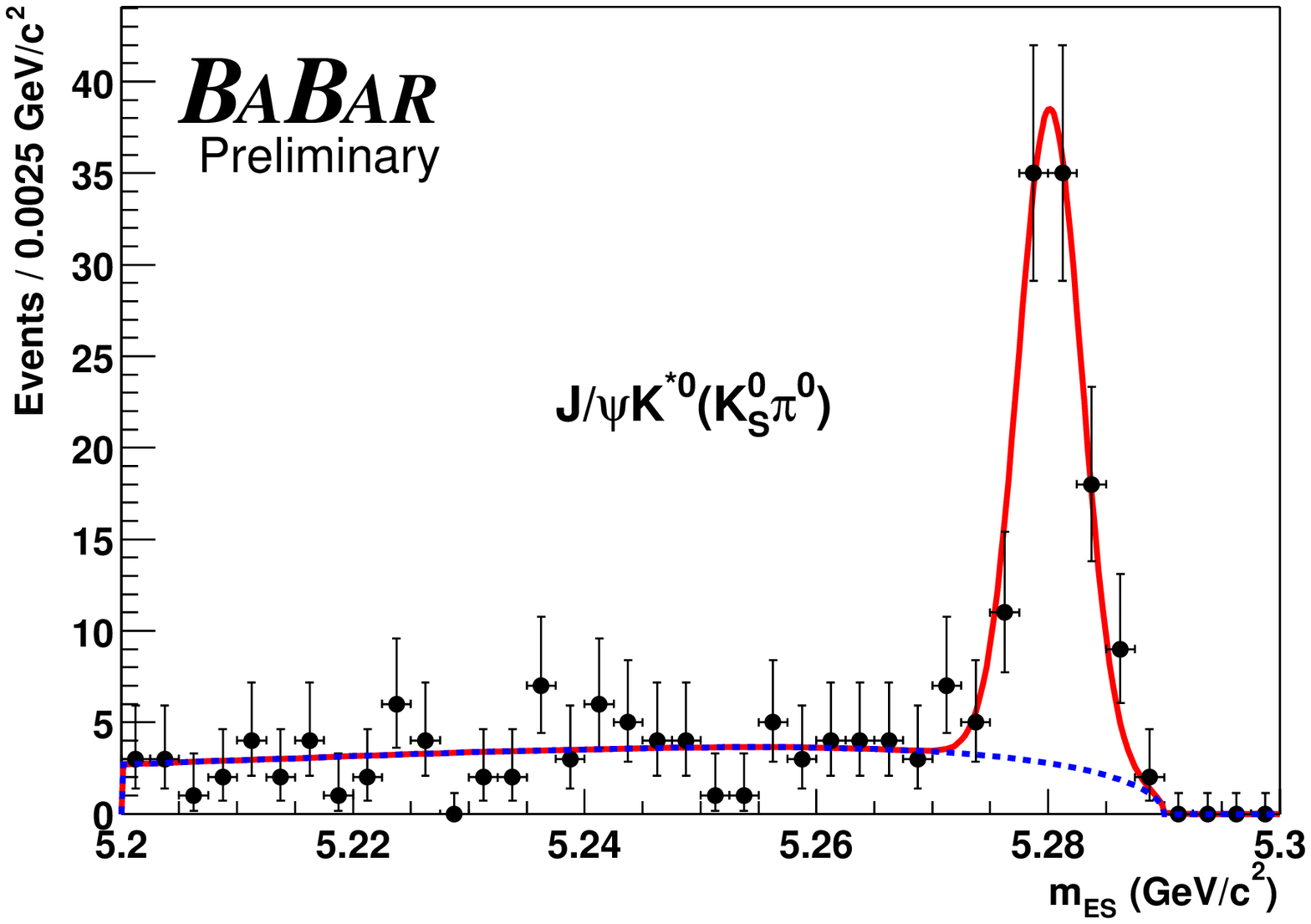,width=0.495\linewidth}
\epsfig{figure=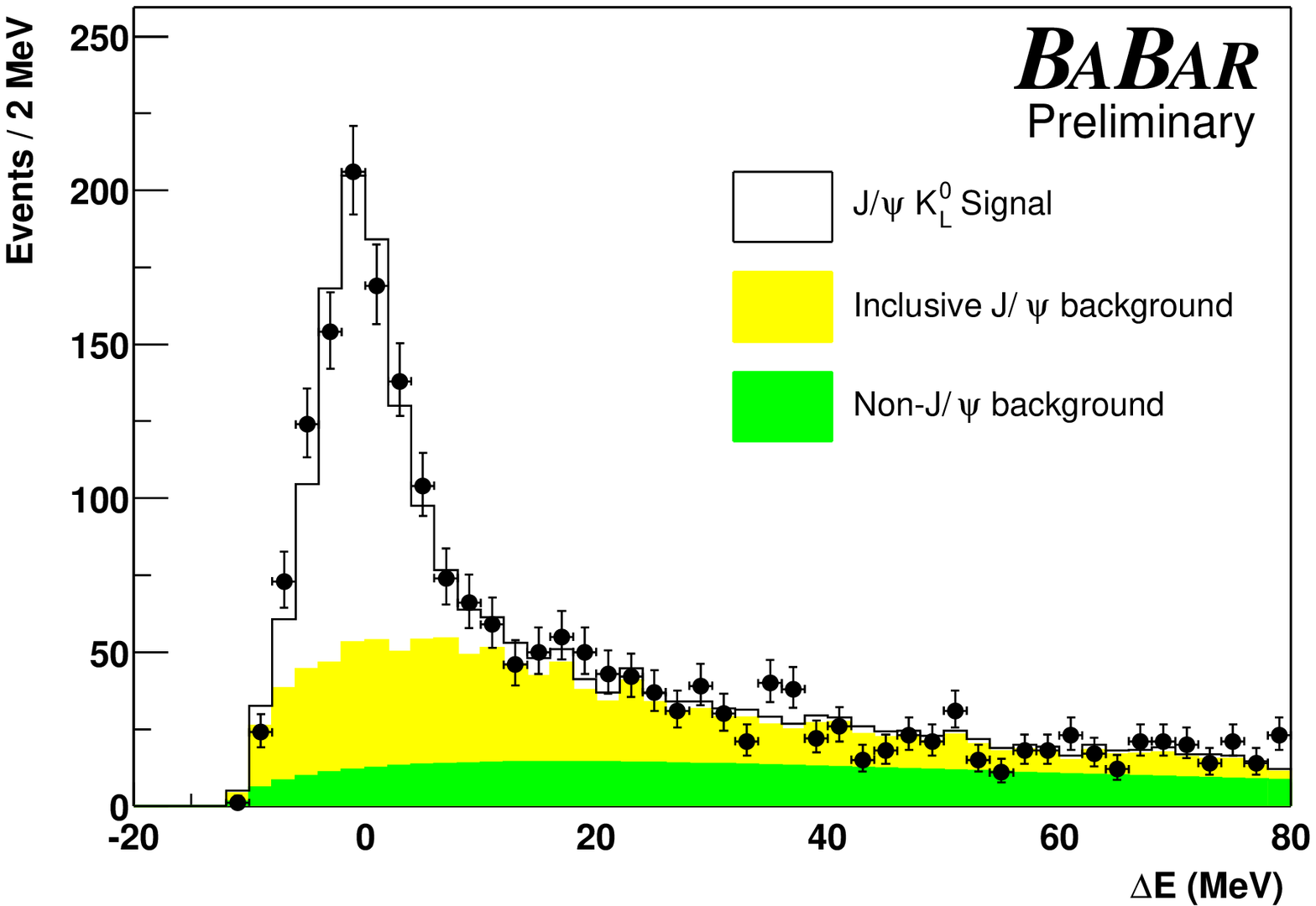,width=0.495\linewidth}
\put(-460,110){{\large e)}}
\put(-215,115){{\large f)}}}
\caption{Distribution of \mes\ for flavor tagged $B_{\CP}$ candidates
  selected in the final states 
  a) $J/\psi K^0_S\ (\KS\to\pip\pim)$, 
  b) $J/\psi K^0_S\ (\KS\to\piz\piz)$, 
  c) $\psi(2S) K^0_S$, 
  d) $\chi_{c1} K^0_S$,
  e) $J/\psi K^{*0}(K^{*0}\to K^0_S\pi^0)$, and
  f) distribution of $\Delta E$ for flavor tagged $\jpsi\KL$
  candidates.}  
\label{fig:bcpsample}
\end{center}
\end{figure}

In Table~\ref{tab:result} we list the numbers of events and the signal 
purities for the tagged $B_{\CP}$ candidates. The purities are
determined from fits to the \mes (all \KS\ modes except $\jpsi\Kstarz$)
or $\Delta E$ (\KL\ mode) distributions in 
data, or from Monte Carlo simulation ($\jpsi\Kstarz$ mode). 
Figure~\ref{fig:bcpsample} shows the \mes distributions for modes
containing a \KS, and $\Delta E$ for the \jpsi\KL
candidates. For modes containing a \KS, 
we use a Monte Carlo simulation to estimate the fractions of
events in the signal peaks that are due to cross-feed from other decay 
modes. The fractions of peaking background
range between $(0.8\pm 0.2)$\% for $\jpsi\KS\ (\KS\to\pip\pim)$ and
$(6.0\pm 1.8)$\% for $\psitwos\KS$. 
For the $\jpsi\KL$ decay mode, the composition, effective $\eta_f$, and
$\Delta E$ distributions of the individual background sources are
determined either from Monte Carlo simulation (for $B$ decays to
\jpsi) or from the $m_{\ell^+ \ell^-}$ sidebands in data (for fake
$\jpsi\to \ell^+ \ell^-$).   
The tagging efficiencies 
for the four tagging categories are measured from data and summarized in 
Table~\ref{tab:mistag}. 
\par
\begin{table}[!thb] 
\caption{ 
Number of tagged events, signal purity, and result of fitting for \CP
asymmetries in the full \CP sample and in various subsamples, as well as
in the $B_{\rm flav}$ and charged $B$ control samples. Purity is the
fitted number of signal events divided by the total number of events in
the $\Delta E$ and \mes signal region defined in the text. Errors are
statistical only.} 
\label{tab:result} 
\begin{ruledtabular} 
\begin{tabular*}{\hsize}{ l@{\extracolsep{0ptplus1fil}} r c@{\extracolsep{0ptplus1fil}} D{,}{\ \pm\ }{-1} } 
 Sample  & $N_{\rm tag}$ & Purity (\%) & \multicolumn{1}{c}{$\ \ \ \stwob$}  
\\ \colrule 
 Full \CP\ sample                        & 1850  & 79 &  0.75, 0.09 \\ 
\hline
$\ \jpsi \KS$ ($\KS \to \pi^+ \pi^-$)    & 693  & 96  &  0.79, 0.11 \\ 
$\ \jpsi \KS$ ($\KS \to \pi^0 \pi^0$)    & 123  & 89  &  0.42, 0.33 \\ 
$\ \psi(2S) \KS$                         & 119  & 89  &  0.84, 0.32 \\
$\ \chicone \KS $                        &  60  & 94  &  0.84, 0.49 \\ 
$\ \jpsi \KL$                            & 742  & 57  &  0.73, 0.19 \\
$\ \jpsi\Kstarz$ ($\Kstarz \to \KS\piz$) & 113  & 83  &  0.62, 0.56 \\ 
\hline
\hline
$\jpsi\KS$, $\psitwos\KS$, $\chicone\KS$ only $(\eta_f=-1)$ &  995  & 94 &  0.76, 0.10 \\ 
\hline 
$\ $ {\tt Lepton} tags                   & 176  & 97  &  0.73, 0.16 \\ 
$\ $ {\tt Kaon} tags                     & 504  & 95  &  0.75, 0.14 \\ 
$\ $ {\tt NT1} tags                      & 117  & 95  &  0.86, 0.33 \\ 
$\ $ {\tt NT2} tags                      & 198  & 94  &  0.84, 0.61 \\ 
\hline 
$\ $ \Bz\ tags                           & 471  & 94  &  0.79, 0.14 \\ 
$\ $ \Bzb\ tags                          & 524  & 95  &  0.73, 0.14 \\ 
\hline\hline
$B_{\rm flav}$ sample                    & 17546 & 85 &  0.00, 0.03 \\
Charged $B$ sample                       & 14768 & 89 & -0.02, 0.03 \\
\end{tabular*} 
\end{ruledtabular} 
\end{table}

\begin{table}[thb] 
\caption
{Average mistag fractions $\mistag_i$ and mistag differences
  $\Delta\mistag_i=\mistag_i(\Bz)-\mistag_i(\Bzb)$, extracted for each
  tagging category $i$ from the maximum-likelihood fit to the time
  distribution for the fully-reconstructed \Bz\ sample ($B_{\rm
  flav}$ and $B_{\CP}$). The figure of merit for tagging is the effective
  tagging efficiency $Q_i = \eps_i (1-2\mistag_i)^2$, where $\eps_i$  is
  the fraction of events with a reconstructed tag vertex that is
  assigned to the $i^{th}$ category. Uncertainties are statistical
  only. The statistical error on \stwob is proportional to $1/\sqrt{Q}$,
  where $Q=\sum Q_i$. }  
\label{tab:mistag} 
\begin{ruledtabular} 
\begin{tabular*}{\hsize}{l
@{\extracolsep{0ptplus1fil}}  D{,}{\ \pm\ }{-1} 
@{\extracolsep{10ptplus1fil}} D{,}{\ \pm\ }{-1} 
@{\extracolsep{0ptplus1fil}}  D{,}{\ \pm\ }{-1} 
@{\extracolsep{0ptplus1fil}}  D{,}{\ \pm\ }{-1}}  
Category     & 
\multicolumn{1}{c}{$\ \ \ \varepsilon$   (\%)} & 
\multicolumn{1}{c}{$\ \ \ \mistag$       (\%)} & 
\multicolumn{1}{c}{$\ \ \ \Delta\mistag$ (\%)} &
\multicolumn{1}{c}{$\ \ \ Q$             (\%)} \\ \colrule  
{\tt Lepton} & 11.1,0.2 &  8.6, 0.9 &  0.6,1.5 &   7.6,0.4  \\  
{\tt Kaon}   & 34.7,0.4 & 18.1, 0.7 & -0.9,1.1 &  14.1,0.6  \\ 
{\tt NT1}    & ~7.7,0.2 & 22.0, 1.5 &  1.4,2.3 &   2.4,0.3  \\ 
{\tt NT2}    & 14.0,0.3 & 37.3, 1.3 & -4.7,1.9 &   0.9,0.2  \\  \colrule 
All          & 67.5,0.5 &           &          &  25.1,0.8  \\ 
\end{tabular*} 
\end{ruledtabular} 
\end{table} 

We determine \stwob with a simultaneous unbinned maximum likelihood fit 
to the \deltat distributions of the $B_{\CP}$ and $B_{\rm flav}$ tagged
samples. Equations~\ref{eq:timedist} (with $|\lambda|=1$)
and~\ref{eq:angle} describe the \deltat\ distribution of the $\eta_f=-
1$ and \jpsi\KL samples, and the \jpsi\Kstarz sample, respectively.
The \deltat distributions of the $B_{\rm flav}$ sample evolve
according to the known frequency for flavor oscillation in neutral $B$
mesons. The observed amplitudes for the \CP asymmetry in the
$B_{\CP}$ sample and for flavor oscillation in the $B_{\rm flav}$ sample 
are reduced by the same factor $(1-2\mistag)$ due to mistags. The
\deltat distributions for the $B_{\CP}$ and $B_{\rm flav}$ samples are
both convolved with a common \deltat resolution function. Events are
assigned signal and background probabilities based on  
the \mes\ (all modes except $\jpsi\KL$) or $\Delta E$ ($\jpsi\KL$)
distributions. Backgrounds are incorporated with an empirical
description of their \deltat evolution, containing prompt and 
non-prompt components convolved with a separate resolution   
function~\cite{babar-stwob-prd}.   
\par
The \deltat\ resolution function ${\cal {R}}$ for the signal  is
represented in terms of $\delta_t \equiv \deltat -\deltat_{\rm true}$ by
a sum of three Gaussian distributions with different means and widths:

\begin{eqnarray}
{\cal {R}}( \delta_{\rm t}) &=&  \sum_{k={\rm core,tail}}
{ \frac{f_k}{S_k\sigma_{\deltat}\sqrt{2\pi}} \, {\rm exp} 
\left(  - \frac{( \delta_{\rm t}-b_k\sigma_{\deltat})^2} 
 {2({S_k\sigma_{\deltat}})^2 }  \right) }  + 
{ \frac{f_{\rm outlier}}{\sigma_{\rm outlier}\sqrt{2\pi}} \, {\rm exp} 
\left(  - { \delta_{\rm t}^2 \over 
 2{\sigma_{\rm outlier}}^2 }  \right) }.
\label{eq:vtxresolfunct}
\end{eqnarray}

\vskip12pt\noindent
For the core and tail Gaussians, we use two separate scale factors  
$S_{\rm core}$ and $S_{\rm tail}$ to multiply the measurement
uncertainty $\sigma_{\deltat}$ that is derived from the vertex fit for
each event. The scale factor for the tail component $S_{\rm tail}$ is
fixed to the value found in Monte Carlo simulation since it is strongly 
correlated with the other resolution function parameters.   
The core and tail Gaussian distributions are allowed to have nonzero 
means to account for any daughters of long-lived charm particles included in 
the $B_{\rm tag}$ vertex. In the resolution function, mean offsets $b_k$
are multiplied by the event-by-event measurement uncertainty
$\sigma_{\deltat}$ to account for an observed correlation between the
mean of the $\delta_{\rm t}$ distribution and the measurement
uncertainty $\sigma_{\deltat}$ in Monte Carlo simulation. The mean of
the core Gaussian is allowed to be different for each tagging
category. One common mean is used for the tail component. The outlier
Gaussian has a fixed width and no offset; it  accounts for the fewer
than $0.5\%$ of events 
with incorrectly reconstructed vertices. 
In simulated events, we find no significant difference between the
$\Delta t$ resolution function of the $B_{\CP}$ sample and of the
$B_{\rm flav}$ sample. This is expected, since the $B_{\rm tag}$ vertex
precision dominates the \deltat\ resolution. Hence, the same resolution
function is used for all modes. 

\begin{table}[!htb] 
\caption{\deltat resolution function parameters for $B_{\rm flav}$ and
  $B_{\CP}$ candidates extracted from the simultaneous
  maximum-likelihood fit to the 
  \deltat distributions for the $B_{\rm flav}$ and $B_{\CP}$ samples.}  
\label{tab:resolution} 
\begin{ruledtabular} 
\begin{tabular*}{\hsize}{lc lc}
$S_{\rm core}$                           & $1.19 \pm 0.07$ & 
$S_{\rm tail}$                          &      3.0 (fixed)  \\
$b_{\rm core}$ ({\tt lepton})       & $ 0.01 \pm 0.07$ &
$b_{\rm tail}$              & $-2.5 \pm 1.7$ \\
$b_{\rm core}$ ({\tt kaon})         & $-0.24 \pm 0.04$ &
$\sigma_{\rm outlier}$                     & 8~ps (fixed) \\
$b_{\rm core}$ ({\tt NT1})           & $-0.20 \pm 0.08$ &
$f_{\rm tail}$                               & $0.05 \pm 0.04$  \\
$b_{\rm core}$ ({\tt NT2})          & $-0.21 \pm 0.06$ &
$f_{\rm outlier}$                           & $0.004 \pm 0.002$ \\
\end{tabular*} 
\end{ruledtabular} 
\end{table} 
\par
A total of 35 parameters are varied in the final fit, including the
values of \stwob (1), the average mistag fraction $\mistag$
and the 
difference $\Delta\mistag$ between \Bz\ and \Bzb\ mistags for each
tagging category (8), parameters for the signal \deltat resolution (8),
and parameters for background time dependence (6), \deltat resolution
(3), and mistag fractions (8). In addition, we allow \ctwob (1), which is
determined from the $\jpsi\Kstarz$ events, to vary in the fit. The sign
of \ctwob cannot be determined due to a twofold ambiguity in the
relative strong phases of the angular amplitudes~\cite{dighedunietz}. We use the
convention for the strong phases given in Appendix~\ref{appendixa}.
\par
The determination of the mistag fractions and \deltat resolution
function for the signal is dominated by the high-statistics $B_{\rm
  flav}$ sample. 
The measured mistag fractions and the parameters of the signal
resolution function are listed in Tables~\ref{tab:mistag} and \ref{tab:resolution}.  
Background parameters are determined from events with
$\mes < 5.27\gevcc$ (except $\jpsi\KL$ and $\jpsi\Kstarz$). We fix
$\tau_{\Bz}=1.548\ps$ and $\deltamd
=0.472\ps^{-1}$~\cite{PDG2000}. The largest correlation between
\stwob\ and any linear combination of the other free parameters is 0.14. 
\par
\begin{figure}[tp]
\begin{center}
\epsfig{figure=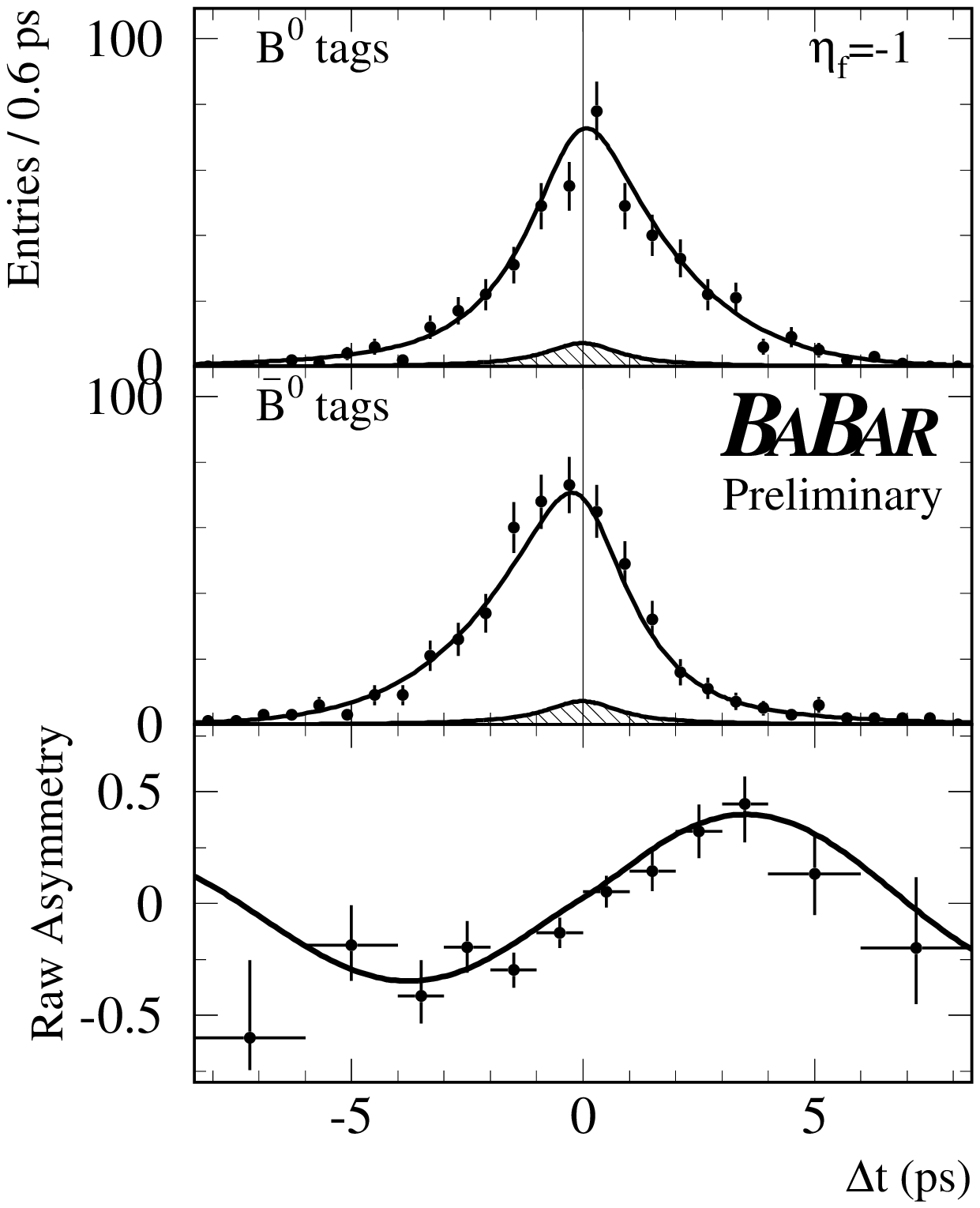,width=0.49\linewidth} 
\epsfig{figure=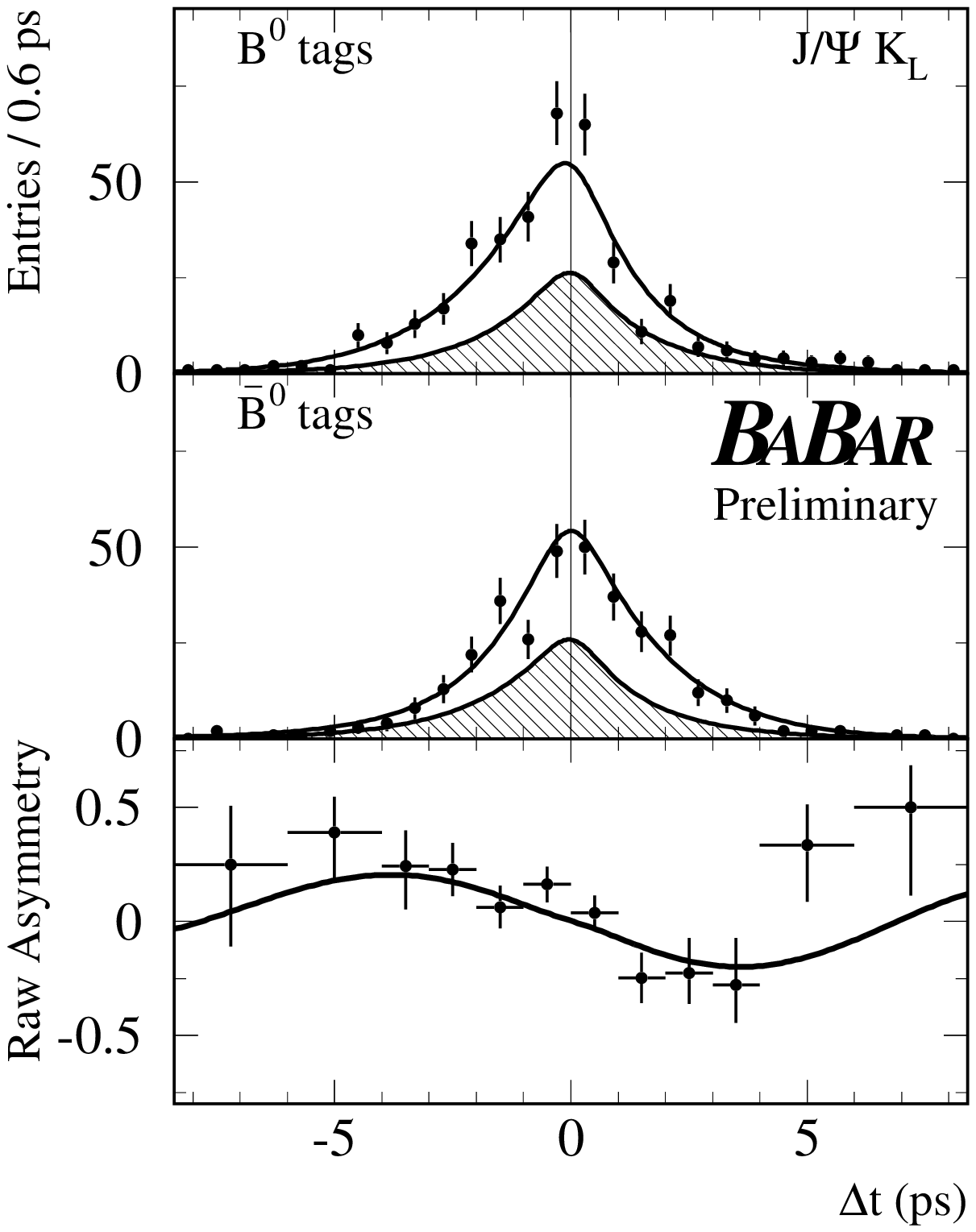,width=0.49\linewidth} 
\put(-435,290){{\large a)}}
\put(-185,290){{\large d)}}
\put(-435,195){{\large b)}}
\put(-185,195){{\large e)}}
\put(-435,115){{\large c)}}
\put(-185,115){{\large f)}}
\caption{Number of $\eta_f=-1$ candidates 
($J/\psi \KS$,
$\psi(2S) \KS$, 
$\chicone \KS$) 
in the signal region a) with a \Bz tag $N_{\Bz }$ and b)
with a \Bzb tag $N_{\Bzb}$, and c) the raw asymmetry
$(N_{\Bz}-N_{\Bzb})/(N_{\Bz}+N_{\Bzb})$, as functions of \deltat . The
solid curves represent the result of the combined fit 
to the full $B_{\rm CP}$ sample.
The shaded regions represent the background contributions.
Figures d) -- f) contain the corresponding information for the $\eta_f=+1$
mode $J/\psi \KL$.
The likelihood is normalized to the total number of \Bz and \Bzb tags. 
The value of \stwob is independent of the individual
normalizations and therefore of the difference between the number of \Bz
and \Bzb tags. This difference is responsible for the small vertical
shift between the data points and the solid curves.}   
\label{fig:cpdeltat}
\end{center}
\end{figure}

The simultaneous fit to all \CP decay modes and the flavor decay modes
yields   
\begin{eqnarray}
\stwob=0.75 \pm 0.09\ \stat \pm 0.04\ \syst.\nonumber
\end{eqnarray}
\noindent
Figure~\ref{fig:cpdeltat} shows the \deltat distributions and 
asymmetries in yields between \Bz tags and \Bzb tags for the
$\eta_f=-1$ and $\eta_f = +1$ samples as a function of \deltat,
overlaid with the projection of the global likelihood fit result. 
\par
Repeating the fit with all parameters except \stwob
fixed to their values at the global maximum likelihood, we attribute a
total contribution in quadrature of $0.01$ to the  
error on \stwob\ due to the combined statistical uncertainties in
mistag rates, \deltat\ resolution, and background parameters. 
The dominant sources of systematic error are 
due to uncertainties in the level, composition, and \CP\ asymmetry of 
the background in the selected \CP events (0.022),  
limited Monte Carlo simulation statistics (0.014), and the assumed
parameterization of the \deltat\ resolution function (0.013), 
due in part to residual uncertainties in the SVT alignment.
Uncertainties in \deltamd and $\tau_{\Bz}$ each contribute 0.010 to the 
systematic error. 
We have performed fits with \deltamd and $\tau_{\Bz}$ fixed to a series
of values around the corresponding world averages in order to determine
the dependence of \stwob on these two parameters and find that $\stwob =
[ 0.75 
- 0.31 (\deltamd - 0.472\ \mbox{ps}^{-1}) 
- 0.62 (\tau_{\Bz} - 1.548\ \mbox{ps})]$.

\par
\begin{figure}[htp]
\begin{center}
\epsfig{figure=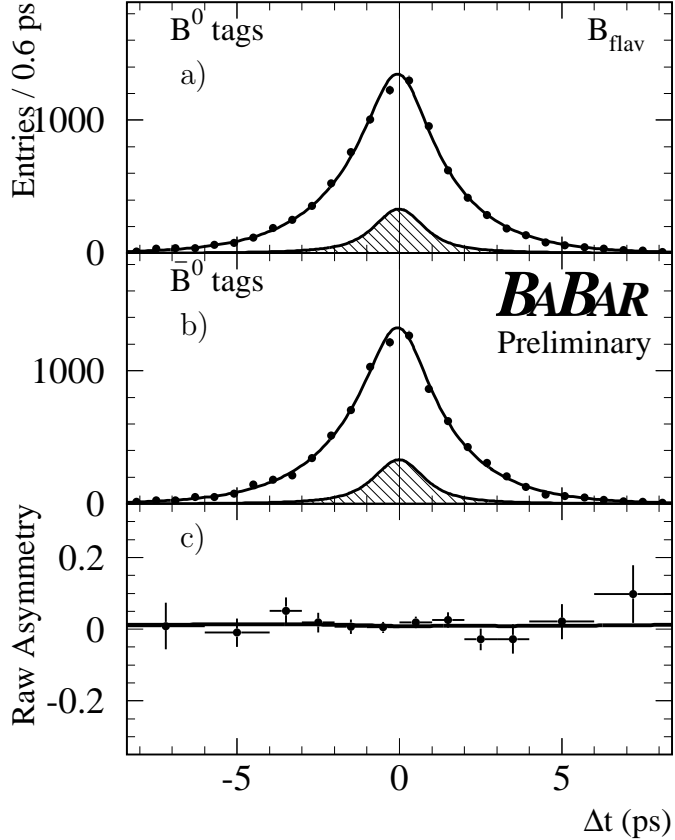,width=0.89\linewidth} 
\put(-310,295){{\large a)}}
\put(-310,200){{\large b)}}
\put(-310,120){{\large c)}}
\caption{Number of $B_{\rm flav}$ candidates 
in the signal region a) with a \Bz tag, $N_{\Bz }$, and b)
with a \Bzb tag, $N_{\Bzb}$, and c) the raw asymmetry
$(N_{\Bz}-N_{\Bzb})/(N_{\Bz}+N_{\Bzb})$, as functions of \deltat . The
solid curves represent the result of the combined fit to all selected
$B_{\rm flav}$ events. The shaded regions represent the background
contributions.}    
\label{fig:bflavasym}
\end{center}
\end{figure}

The large sample of reconstructed events allows a number of consistency 
checks, including separation of the data by decay mode, tagging category,
and $B_{\rm tag}$ flavor. The results of fits 
to these subsamples for the $\eta_f=-1$ sample are shown in
Table~\ref{tab:result} and found to be statistically consistent.
The fit results to the samples of non-\CP decay modes
indicate no statistically significant asymmetry. The distributions and 
asymmetry in yields for \Bz and 
\Bzb tags as a function of \deltat for the $B_{\rm flav}$ sample are
shown in Fig.~\ref{fig:bflavasym}. In addition,
we have made a number of detailed analyses of the expected distribution
of changes in \stwob that might result from reprocessing, in order to
account for the correlations between the two results from the same
sample. From these studies, we conclude that the observed difference in
the 1999--2000 result, before and after reprocessing, is equivalent to
about two standard deviations of the distribution of predicted changes
due to reprocessing for events that appear in common. The change
in the overall result in the 1999--2000 dataset, from $\stwob=0.45\pm 0.20$ to
$0.60\pm 0.15$, is consistent with the effects of both reprocessing and
event selection modifications. 
\par
With the theoretically preferred choice of the strong phases, 
consistent with the hypothesis of the $s$-quark helicity conservation in
the decay~\cite{suzuki}, the parameter $\ctwob$ is measured to
be $+3.3 ^{+0.6}_{-1.0}\ \stat ^{+0.6}_{-0.7}\ \syst$. This value is 2.2
$\sigma$ away from the one obtained using the relation $\sqrt{1-\sin^2
  2\beta}=0.66$.  
The dominant contributions to the systematic error on \ctwob are due
to uncertainties in the transversity amplitudes for the signal
($_{+0.2}^{-0.4}$) and the background ($\pm 0.5$).   
If we fix $\ctwob$ to 0.66, the measured value of
\stwob does not change. For the alternative set of strong phases,
$(\phi_\perp,\ \phi_\parallel)\ \to\ (\pi-\phi_\perp,\
-\phi_\parallel)$, the sign of \ctwob flips, yielding $\ctwob = -3.3
^{+1.0}_{-0.6}\ \stat ^{+0.7}_{-0.6}\ \syst$. 
\par
If the parameter $\vert\lambda\vert$ in Eq.~\ref{eq:timedist} is
allowed to float in the fit to the $\eta_f=-1$ sample, which has high
purity and requires minimal assumptions on the effect of backgrounds,
the value obtained is $\vert\lambda\vert = 0.92 \pm 0.06\ \stat \pm
0.02\ \syst$. The sources of the systematic error are the same as for
the \stwob measurement with an additional contribution in quadrature of
0.012 from the uncertainty on the difference in the tagging
efficiencies for \Bz and \Bzb tagged events. In this fit, the coefficient
of the $\sin(\deltamd \deltat)$ term in Eq.~\ref{eq:timedist} is
measured to be $0.76\pm 0.10\ \stat$ in agreement with
Table~\ref{tab:result}. 

This analysis supersedes our previous result~\cite{babar-stwob-prl}. It
provides the single most precise measurement of \stwob currently
available and is consistent with the range implied by measurements and
theoretical estimates of the magnitudes of CKM matrix 
elements in the context of the Standard Model~\cite{CKMconstraints}.  

We are grateful for the 
extraordinary contributions of our \pep2\ colleagues in
achieving the excellent luminosity and machine conditions
that have made this work possible.
The success of this project also relies critically on the 
expertise and dedication of the computing organizations that 
support \babar.
The collaborating institutions wish to thank 
SLAC for its support and the kind hospitality extended to them. 
This work is supported by the
US Department of Energy
and National Science Foundation, the
Natural Sciences and Engineering Research Council (Canada),
Institute of High Energy Physics (China), the
Commissariat \`a l'Energie Atomique and
Institut National de Physique Nucl\'eaire et de Physique des Particules
(France), the
Bundesministerium f\"ur Bildung und Forschung
(Germany), the
Istituto Nazionale di Fisica Nucleare (Italy),
the Research Council of Norway, the
Ministry of Science and Technology of the Russian Federation, and the
Particle Physics and Astronomy Research Council (United Kingdom). 
Individuals have received support from 
the A. P. Sloan Foundation, 
the Research Corporation,
and the Alexander von Humboldt Foundation.

\vskip12pt 
 
\appendix

\section{Time-dependent \CP asymmetry in $B\to J/\psi K^{*0} (K^{*0}\to
  \KS\pi^0$)}\label{appendixa}
\begin{figure}[!hbtp]
\begin{center}
\epsfxsize8.6cm
\figurebox{}{\textwidth}{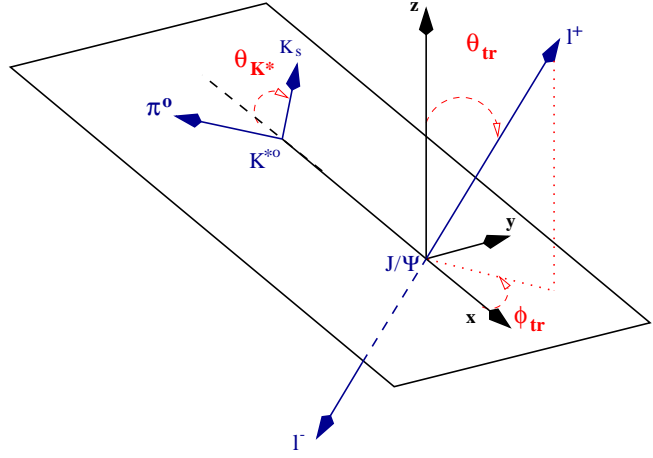}
\caption{Definitions of transversity angles \thetatr , \phitr, and
  \thetakstar . The angles \thetatr and \phitr  are determined in the
  \jpsi rest frame. The angle  \thetakstar is determined in the \Kstar
  rest frame.}
\label{fig:jpsiksttrans}
\end{center}
\end{figure}

The decay $B \to \jpsi \Kstar$ is described by three  
amplitudes. In the transversity basis \cite{ref:dunietz,ref:dighe}, the
amplitudes \az, \ap\ and \at\ have \CP\ eigenvalues $+1, +1$ and $-1$,
respectively. 
\az\ corresponds to longitudinal polarization of the vector mesons, and
\ap\ and \at\ correspond to parallel and perpendicular transverse
polarizations, respectively. 
The relative phase between the parallel (perpendicular)
transverse amplitude and the longitudinal amplitude is given by 
$\phi_\parallel\equiv\mbox{arg}(\ap/\az )$ 
($\phi_\perp\equiv\mbox{arg}(\at/\az )$).
The transversity frame is defined
as the \jpsi\ rest frame (see Fig.~\ref{fig:jpsiksttrans}). The \Kstar\
direction defines the negative $x$ 
axis. The $K\pi$ decay plane defines the $(x,y)$ plane, with $y$ oriented such
that $p_y(K) > 0$. The $z$ axis is the normal to this plane, and the
coordinate system is right-handed.  The transversity angles 
\thetatr and \phitr are defined as the polar and azimuthal angles of the
positive lepton from the \jpsi decay; \thetakstar\ is the \Kstar\ helicity
angle defined in the \Kstar\ rest frame as the angle between the $K$
direction and the direction opposite to the \jpsi. 

\par

The time- and transversity-angle-dependent decay rate distributions
${\rm f}_{+} ({\rm f}_{-})$ when the tagging meson is a $\Bz (\Bzb)$ are
given by 
\begin{eqnarray}
{\rm f}_{\pm}(\deltat,\vec{\omega}) &=& 
\frac{e^{-|\deltat|/\tau_{\Bz}}}{4\tau_{\Bz}} \bigg[ 
{I(\vec{\omega};\vec{\cal A})} \mp  \Big\{
{C(\vec{\omega};\vec{\cal A})}
\cos(\deltamd \deltat) \nonumber \\
& & +\Big[
{S_{\stwob}{(\vec{\omega};\vec{\cal A})}}\stwob  + 
{S_{\ctwob}{(\vec{\omega};\vec{\cal A})}}\ctwob \Big]
\sin(\deltamd \deltat) \Big\}\bigg].  
\end{eqnarray}
The coefficients $I$, $C$, $S_{\stwob}$, and $S_{\ctwob}$, which 
depend on the transversity angles $\vec{\omega} =
(\theta_{K^*},\thetatr,\phitr)$ and the transversity amplitudes
$\vec{\cal A} = ({A_0,A_\parallel,A_\perp})$, are given by
\begin{eqnarray}
I(\vec{\omega};\vec{\cal A}) & = & f_1(\vec{\omega}) |A_0 |^2 +
f_2(\vec{\omega}) |A_\parallel|^2 + f_3(\vec{\omega}) |A_\perp|^2 +
f_5(\vec{\omega}) |A_\parallel| |A_0| \cos(\phi_\parallel -
\phi_0) \nonumber\\ 
C(\vec{\omega};\vec{\cal A}) & = & f_4(\vec{\omega}) |A_\parallel
||A_\perp|\sin(\phi_\perp - \phi_\parallel)  + f_6(\vec{\omega})
|A_\perp| |A_0 |\sin(\phi_\perp - \phi_0) \nonumber\\ 
S_{\stwob}(\vec{\omega};\vec{\cal A}) & = & f_1(\vec{\omega}) |A_0 |^2 +
f_2(\vec{\omega}) |A_\parallel|^2 - f_3(\vec{\omega}) |A_\perp|^2 +
f_5(\vec{\omega}) |A_\parallel| |A_0| \cos(\phi_\parallel - \phi_0)
\nonumber \\ 
S_{\ctwob}(\vec{\omega};\vec{\cal A}) & = & -f_4(\vec{\omega})
|A_\parallel ||A_\perp| \cos(\phi_\perp - \phi_\parallel) -
f_6(\vec{\omega}) |A_\perp| |A_0 |\cos(\phi_\perp - \phi_0)
\end{eqnarray}
with 
\begin{eqnarray}
f_1(\vec{\omega}) & = & \frac{9}{32\pi} 2  \cos^{2} (\theta_{K^{*}})
\left[ 1 - \sin^{2}(\theta_{tr})\cos^{2}(\varphi_{tr}) \right]\nonumber \\ 
f_2(\vec{\omega}) & = & \frac{9}{32\pi}\sin^{2} (\theta_{K^{*}}) \left[1
  - \sin^{2}(\theta_{tr})\sin^{2}(\varphi_{tr}) \right] \nonumber \\ 
f_3(\vec{\omega}) & = & \frac{9}{32\pi}\sin^{2}
(\theta_{K^{*}})\sin^{2}(\theta_{tr}) \nonumber \\ 
f_4(\vec{\omega}) & = & \frac{9}{32\pi}\sin^{2} (\theta_{K^{*}})
\sin(2\theta_{tr})\sin(\varphi_{tr}) \nonumber \\ 
f_5(\vec{\omega}) & = & - \frac{9}{32\pi}\frac{1}{\sqrt{2}} \sin(2
\theta_{K^{*}})  \sin^{2}(\theta_{tr})\sin(2\varphi_{tr}) \nonumber \\ 
f_6(\vec{\omega}) & = & \frac{9}{32\pi}\frac{1}{\sqrt{2}} \sin(2
\theta_{K^{*}}) \sin(2\theta_{tr})\cos(\varphi_{tr}). 
\end{eqnarray}

\end{document}